\documentclass[aps,twocolumn,superscriptaddress,amssymb,amsmath,prb,showpacs]{revtex4}

\usepackage{graphicx}

\newcommand{\mrm}{\mathrm}
\newcommand{\mbf}{\mathbf}
\newcommand{\GL}{\Gamma_\mathrm{L}}
\newcommand{\GR}{\Gamma_\mathrm{R}}

\begin{document}

\title{Generation of Spin Entanglement in Nonequilibrium Quantum Dots}

\author{Stefan Legel}
\affiliation{Institut f\"ur Theoretische Festk\"orperphysik and DFG-Center for Functional Nanostructures (CFN), Universit\"at Karlsruhe, 76128 Karlsruhe, Germany}

\author{J\"urgen K\"onig}
\affiliation{Institut f\"ur Theoretische Physik III, Ruhr-Universit\"at Bochum, 44780 Bochum, Germany}

\author{Guido Burkard}
\affiliation{Department of Physics and Astronomy, University of Basel, Klingelbergstrasse 82, CH-4056 Basel, Switzerland}
\affiliation{Institut f\"ur Theoretische Physik C, RWTH Aachen,
D-52056 Aachen, Germany}

\author{Gerd Sch\"on}
\affiliation{Institut f\"ur Theoretische Festk\"orperphysik and DFG-Center for Functional Nanostructures (CFN), Universit\"at Karlsruhe, 76128 Karlsruhe, Germany}

\date{\today}

\begin{abstract}
We propose schemes for generating spatially-separated spin entanglement 
in systems of two quantum dots with onsite Coulomb repulsion 
weakly coupled to a joint electron reservoir. 
An enhanced probability for the formation of spin entanglement is found
in nonequilibrium situations with one extra electron on each dot, 
either in the transient state after rapid changes of the gate voltage, or
in the steady state with applied bias voltage.
In both cases so-called Werner states with with spin singlet fidelity 
exceeding $1/2$ are generated, which indicates entanglement.
\end{abstract}

\pacs{03.67.Mn,73.23.Hk,73.21.La,73.63.Kv}

\maketitle

\section{Introduction}

The entanglement of quantum states is one of the cornerstones of quantum
information processing.~\cite{BennettDiVincenzo} 
Entangled photons have been used in experiments in quantum communication and 
cryptography.~\cite{Gisin} 
For electrons in a solid-state environment recent progress has been linked to 
advances in fabrication technology for nano-scale 
devices.~\cite{PettaScience,KoppensNature}
The availability of an electron spin entangler in a solid-state environment  
would allow the implementation of quantum communication schemes with electron 
spins.~\cite{BLS,BL}
Several schemes have been suggested for the production of spatially-separated
entangled electrons in solid state systems.
Many of them rely on extracting the entangled electrons of a Cooper pair from
a superconductor and separate them into two normal leads,~\cite{Lesovik01}
Luttinger liquids,~\cite{luttinger} or to two leads through two quantum 
dots.~\cite{LossPRB63}
Others are based on separating the electrons forming a spin singlet on a 
double-quantum dot,~\cite{blaauboer} using interference effects in a quantum
dot in the cotunneling regime,~\cite{OliverPRL88} separating a pair of
entangled electrons from a singlet state by a triple quantum 
dot,~\cite{SaragaLoss} or scattering off magnetic impurities.~\cite{scattering}

\begin{figure}
\begin{center}
\includegraphics[height=3.4cm]{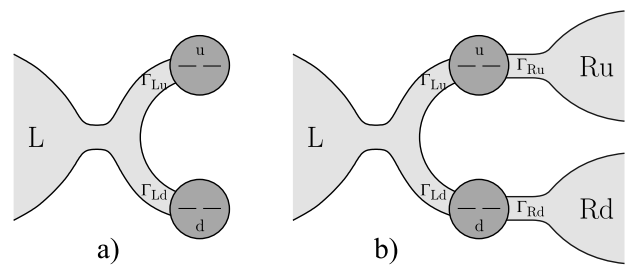}
\end{center}
\caption{ \label{setups} The setups: 
a) Two quantum dots (u and d) are coupled to a joint electron reservoir (L).
b) In addition to a), the quantum dots are coupled to two independent 
reservoirs (Ru and Rd) on the right.}
\end{figure}

In this article we show that a pair of entangled electrons can be created by
driving out of equilibrium a system of two quantum dots with onsite
Coulomb repulsion and weak coupling to a joint electron reservoir.
Specifically we consider the two setups depicted in Fig.~\ref{setups}.
Electrons enter the dots from the reservoirs, and we consider the nonequilibrium state with one electron on each dot. In setup a) we study the transient behavior after quickly pushing the dot levels from above to below the Fermi energy of the lead, and find an enhanced
probability for the singlet state as compared to a triplet. In setup b) we drive the system out of equilibrium by applying a bias voltage between left and right leads. Depending on the polarity of the applied bias, we find in the steady state an enhanced probability of either the singlet or the triplet states. 
The mixed states with two electrons in the two dots represent so-called Werner states.~\cite{Werner} 
In the case where the electrons entered from the common (left) reservoir we find 
regimes where the Werner fidelity is larger than $1/2$, which implies a high probability for the formation of a singlet state.

\section{Model}

The Hamiltonian of the system is 
\begin{equation}
  H = H_\mrm{dots} + H_\mrm{leads} + H_\mrm{tunnel}.
\end{equation}
The two quantum dots, $i=\mrm{u},\mrm{d}$ (up and down), described by
\begin{equation} 
  H_\mrm{dots}= \sum_{i} \left[
    \sum_\sigma \varepsilon_i \, c_{i\sigma}^\dagger c_{i\sigma} + U \,
    c_{i\uparrow}^\dagger c_{i\downarrow}^\dagger c_{i\downarrow}
    c_{i\uparrow}\right] \, , 
\end{equation}
contain each a single, spin-degenerate energy level $\varepsilon_i$.
In general the dot levels are detuned by $\Delta \varepsilon = \varepsilon_\mrm{u} - \varepsilon_\mrm{d}$.
We assume strong Coulomb repulsion within each dot,
$U\gg k_{\rm B}T, {\rm e}V, \Gamma$
which suppresses double occupancy of each dot.
(Our analysis can be generalized to finite interdot charging energy, which does not change the conclusions qualitatively.)
The leads
\begin{equation}
  H_\mrm{leads}= \sum_r \sum_{k\sigma} \varepsilon_{rk} \, a_{rk\sigma}^\dagger  a_{rk\sigma}\, , 
\end{equation}
with $r=\mrm{L},\mrm{R_u},\mrm{R_d}$, serve as equilibrium reservoirs with electrochemical potentials $\mu_r$.
The tunneling between leads and dots is modeled by 
\begin{equation}
  H_\mrm{tunnel}= \sum_{r} \sum_{k\sigma i} \left( t_{ri} \,
  c^\dagger_{i\sigma} a_{rk\sigma} + \mrm{h.c.} \right)\, .
\end{equation}
The tunneling strength of quantum dot $i$ to reservoir $r$ is parametrized by 
$\Gamma_{ri} = 2\pi\, t_{ri}^2\,N_r$, where $N_r$ denotes the reservoir density of 
states, and we chose a gauge in which all the tunnel
amplitudes $t_{ri}$ are real.
We further define $\Gamma_r \equiv \sum_i \Gamma_{ri}/2$.
For setup a) we have $\Gamma_\mrm{R_u}=\Gamma_\mrm{R_d}=0$.

Our proposal is based on the observation that the
states in the common left lead are only coupled to a certain linear combination of the two quantum dot states.
If Coulomb interaction were absent, $U=0$, filling the double dot with two 
electrons with opposite spin from the common reservoir would lead to the 
product state 
$\left( t_{{\rm Lu}} c^\dagger_{\mrm{u}\sigma} + t_{{\rm Ld}} 
c^\dagger_{\mrm{d}\sigma} \right) \left( t_{{\rm Lu}} 
c^\dagger_{\mrm{u}\bar \sigma} + t_{{\rm Ld}} c^\dagger_{\mrm{d}\bar \sigma} 
\right) |0\rangle =  t_{{\rm Lu}}^2 |\sigma \bar \sigma, 0\rangle +   
t_{{\rm Lu}} t_{{\rm Ld}} \left( |\sigma, \bar \sigma \rangle - 
|\bar \sigma, \sigma \rangle \right) + t_{{\rm Ld}}^2 
|0,\sigma \bar \sigma \rangle$.
For strong Coulomb repulsion, however, the parts that involve double 
occupancy of either dot are projected out, and the final state is, 
$|\sigma, \bar \sigma \rangle - |\bar \sigma, \sigma \rangle$, no product state
but a spin singlet.
No triplet component, although energetically degenerate to the singlet, 
is generated.

In realistic situations various mechanisms will relax the imbalance between the population of spin singlet and triplet states, e.g., tunnel coupling to the right reservoirs shown in Fig.~\ref{setups}b), or a finite detuning $\Delta \varepsilon$. Furthermore, a coupling to an external bath which mediates spin-flip processes or creates a phase difference between the dot states causes an equilibration between singlet and triplet.
In this article, we study in detail nonequilibrium scenarios characterized by the competition between the creation of singlet and triplet states and the relaxation.

\section{Kinetic equations}

For this purpose we employ the real-time diagrammatic technique developed for single quantum dots~\cite{Real-timeDiagrammatic} and extended to multi-dot systems.~\cite{double-dots,Pohjola} 
In this technique the electronic degrees of freedom of the 
leads are integrated out, which results in an effective description in terms of the degrees of freedom of the dot subsystem only.
The dynamics of the latter is then described by a reduced density matrix
with elements $p_{\chi}^{\chi'} \equiv \Big\langle |\chi\rangle 
\langle\chi' | \Big\rangle$, where $\chi$ and $\chi'$ label the
double-dot states, and $\langle \ldots \rangle$ denotes quantum statistical 
expectation values.
In the present case, the Hilbert space of the quantum-dot degrees of freedom 
is spanned by 9 basis states $|\chi_\mrm{u},\chi_\mrm{d}\rangle$, with 
$\chi_i \in \{ 0,\uparrow,\downarrow \}$ denoting the occupation of dot $i$.

The time evolution of the reduced density matrix in the Markovian limit is governed by the kinetic equations~\cite{Real-timeDiagrammatic} 
\begin{equation}
\label{master}
  \frac{d}{d t} p_\chi^{\chi'} + i ( E_{\chi'} - E_\chi) p_\chi^{\chi'}
  =
  \sum_{\chi'' \chi'''} W_{\chi \chi''}^{\chi' \chi'''}
  \, p_{\chi''}^{\chi'''} \, .
\end{equation}
The energy difference $E_{\chi'} - E_\chi$ between states $\chi'$ and $\chi$
leads to a time-dependent phase of the off-diagonal matrix elements.
Transitions due to the tunnel coupling to the leads are described by the
kernels $W_{\chi \chi''}^{\chi' \chi'''}$, the general form of which are given 
in Refs.~\onlinecite{Real-timeDiagrammatic,double-dots}.
In the following we restrict our attention to the limit of weak coupling
and small detuning $\Delta \varepsilon$, where it is sufficient to evaluate 
the kernels in first order in the tunneling strength $\Gamma_{ri}$ and to 
zeroth order in $\Delta \varepsilon$.

To proceed it is convenient to switch to a basis $\{|\chi\rangle \}$ which reflects the symmetries of the problem.
One of the basis states is the empty-dots state $|0\rangle \equiv |0,0\rangle$. For two electrons, one in each dot, the natural basis states are the spin singlet 
$|\mrm{S} \rangle \equiv (|\uparrow, \downarrow\rangle - 
|\downarrow, \uparrow\rangle)/\sqrt{2}$ and triplet states 
$|\mrm{T}_+\rangle \equiv |\uparrow, \uparrow\rangle $, $|\mrm{T}_0 \rangle
\equiv (|\uparrow, \downarrow\rangle + |\downarrow,
\uparrow\rangle)/\sqrt{2}$, and $|\mrm{T}_- \rangle \equiv |\downarrow,
\downarrow\rangle$.
The states with one electron in the double dot can be characterized by the
physical spin $\sigma$ of the electron, as well as by an isospin defined in the
2-dimensional Hilbert space spanned by the two orbital dot levels.
One natural quantization axis ${\mbf n}$ for the isospin operator
$\mbf{I}_\sigma$ is
the one in which the eigenstates of $\mbf{I}_\sigma \cdot \mbf{n}$ are
$| + \rangle_{\mbf{I}_\sigma \cdot \mbf{n}} \equiv |\sigma,0\rangle$ and 
$| - \rangle_{\mbf{I}_\sigma \cdot \mbf{n}} \equiv |0,\sigma \rangle$, 
corresponding to the electron in dot u and d, respectively.
This is motivated by the observation that both the Coulomb interaction 
 and the detuning $\Delta \varepsilon$ in the Hamiltonian is 
diagonal in this isospin basis.
An alternative choice is the axis ${\mbf m}$ defined by 
$| + \rangle_{\mbf{I}_\sigma \cdot \mbf{m}} \equiv \left( 
t_\mrm{Lu}|\sigma,0\rangle + t_\mrm{Ld}|0,\sigma \rangle \right)
/ \sqrt{ t_\mrm{Lu}^2 + t_\mrm{Ld}^2}$ and 
$| - \rangle_{\mbf{I}_\sigma \cdot \mbf{m}} \equiv \left( 
t_\mrm{Ld}|\sigma,0\rangle - t_\mrm{Lu}|0,\sigma \rangle \right)
/ \sqrt{ t_\mrm{Lu}^2 + t_\mrm{Ld}^2}$.
This reflects the fact that filling the double dot by tunneling with one electron 
from the left lead generates the isospin component
$| + \rangle_{\mbf{I}_\sigma \cdot \mbf{m}}$ only.~\cite{Groth}
In this sense, the left lead can, in analogy to magnetism, be viewed as a 
fully isospin-polarized lead with only $+$ isospin-electron states available.
The right reservoirs in setup b), on the other hand, correspond to an 
isospin-unpolarized lead.
In general, the two axes ${\mbf n}$ and ${\mbf m}$ are not orthogonal,
except for the symmetric case when the tunneling strengths to dot u and d are equal,
as can be seen from ${\mbf n} \cdot {\mbf m} = (\Gamma_\mrm{Lu} - 
\Gamma_\mrm{Ld})/(\Gamma_\mrm{Lu} + \Gamma_\mrm{Ld})$.

The total Hamiltonian is invariant under
rotations in spin space, i.e., spin is a conserved quantum number.
Spin symmetry implies $\langle \mbf{I}_\uparrow \rangle
= \langle \mbf{I}_\downarrow \rangle \equiv {\mbf I}/2$
as well as $p_{\mrm{T}_-} = p_{\mrm{T}_0} = p_{\mrm{T}_+} \equiv p_{\rm T}/3$,
which reduces the number of independent 
density matrix elements.
Those are the isospin $\mbf I$ and $\mbf{p} = \left( p_0 , p_1 , p_\mrm{S} , 
p_\mrm{T} \right)$, where $p_1 \equiv \sum_{i\sigma} p_{i\sigma}$ is the 
probability for single occupation.
In this representation, the kinetic equations read
\begin{widetext}
\begin{eqnarray}
\frac{d}{dt} \mbf{p} &=& \sum_{r = \mrm{L},\mrm{R}}\Gamma_r \left(
\begin{matrix}
  -4\,f_r  & 1-f_r & 0 & 0 \\
  4\,f_r & -1-f_r & 2 - 2 f_r & 2 - 2 f_r \\
  0 & f_r/2 & -2 + 2 f_r & 0 \\
  0 & 3 f_r/2 & 0 & -2 + 2 f_r
\end{matrix}
\right) \mbf{p}
+
\Gamma_\mrm{L} \left(
\begin{matrix}
  2 - 2 f_\mrm{L} \\
  -2 + 4 f_\mrm{L} \\
  f_\mrm{L} \\ 
  -3 f_\mrm{L}
\end{matrix}
\right) (\mbf{I} \cdot \mbf{m})
+
2 \Gamma_\mrm{L} f_\mrm{L}\left(
\begin{matrix}
  0 \\
  1 \\
  -1\\
  0 \\
\end{matrix}
\right) (\mbf{I} \cdot \mbf{n}) (\mbf{m} \cdot \mbf{n})
\nonumber \\ \nonumber \\ 
\frac{d}{dt} \mbf{I} &=& \Gamma_\mrm{L} \left[
  2 f_\mrm{L} p_0 + \left( f_\mrm{L} - \frac{1}{2} \right) p_1 + 
  (1 - f_\mrm{L}) p_\mrm{S} - (1 - f_\mrm{L}) p_\mrm{T} \right] \mbf{m}
+
\Gamma_\mrm{L} \left[
  \frac{f_\mrm{L}}{2} p_1 -2 (1 - f_\mrm{L})p_\mrm{S} \right]
\mbf{n} (\mbf{m}\cdot\mbf{n}) 
\nonumber \\
&&
- \sum_{r=\mrm{L},\mrm{R}} \Gamma_r \left( 1 + f_r \right) \mbf{I}
+ \Delta \tilde \varepsilon (\mbf{n} \times \mbf{I}) \, ,
\label{eq:kinetic}
\end{eqnarray}
\end{widetext}
where $f_r = \left[ 1 + \exp (\beta(\varepsilon-\mu_r)) \right]^{-1}$ 
is the Fermi distribution of the electrons in lead $r$.
Here we introduced, apart from the detuning $\Delta \varepsilon$ 
also the average $\varepsilon = (\varepsilon_\mrm{u} + \varepsilon_\mrm{d})/2$ of the dot energies. 
The level detuning is renormalized by the tunneling and given by
$\Delta \tilde \varepsilon = \Delta \varepsilon - 
\frac{\Gamma_\mrm{Lu} - \Gamma_\mrm{Ld}}{2\pi} \left[
\ln\left(\frac{\beta D}{2\pi}\right) - \mrm{Re} \Psi \left( \frac{1}{2} + i
\frac{\beta(\varepsilon-\mu_\mrm{L})}{2\pi} \right) \right]
$,
where $D$ is an high-energy cutoff provided by either Coulomb interaction $U$
or bandwidth of the leads.

\section{Results}

\subsection{Spin Entanglement in Transient States}

\begin{figure}
\begin{center}
\includegraphics[width=8.2cm]{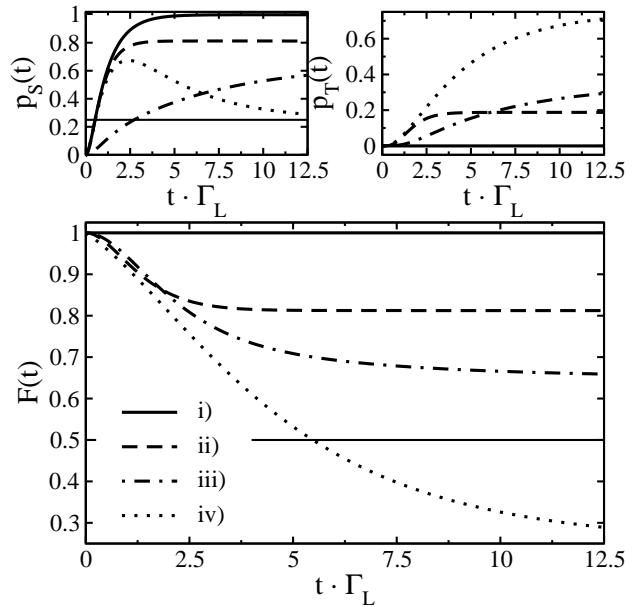}
\end{center}
\caption{\label{DiscForkPlots} 
Upper panels: time evolution of the probabilities for a singlet and a triplet 
state.
Lower panel: the corresponding Werner fidelity. 
For the perfectly symmetric setup, $\Delta\varepsilon = 0$, 
$\Gamma_{\mrm{Lu}} = \Gamma_{\mrm{Ld}}$, in the absence of spin relaxation,
$\Gamma_{\mrm{S}\rightarrow\mrm{T}} = 0$, curve i), we find $F \equiv 1$.
The Werner fidelity is reduced for either ii) nondegenerate dot energy levels, 
$\Delta\varepsilon = \GL$, iii) asymmetric coupling 
$\Gamma_{\mrm{Ld}} = 0.1\,\Gamma_{\mrm{Lu}}$, 
iv) a finite spin relaxation rate $\Gamma_{\mrm{S}\rightarrow\mrm{T}} = 0.2\,\GL$. 
The high-energy cutoff is set to  $D = 100\,\mrm{k_B}T$.}
\end{figure}

Inspection of Eqs.~\eqref{eq:kinetic} reveals
how an imbalance of singlet and triplet states can occur.
When filling the empty double dot with one electron, a finite isospin
along $\mbf m$ is generated.
This in turn, blocks the generation of triplet states as opposed to singlet 
states when filling the double dot with a second electron.
This mechanism becomes most transparent for 
$\Gamma_\mrm{Lu} = \Gamma_\mrm{Ld}$, $\Gamma_\mrm{R_u}=\Gamma_\mrm{R_d}=0$ and 
$\Delta \varepsilon =0$.
In this case, the two equations
\begin{equation}
  \frac{d}{dt} p_\mrm{T} = 
  3\Gamma_\mrm{L}f_\mrm{L} \left( \frac{p_1}{2}-\mbf{I}\cdot\mbf{m} \right)
  -2\Gamma_\mrm{L}(1-f_\mrm{L}) p_\mrm{T}
\end{equation}
\begin{equation}
  \frac{d}{dt} \left( \frac{p_1}{2}-\mbf{I}\cdot\mbf{m} \right) =
  - 3 \Gamma_\mrm{L}f_\mrm{L} \left( \frac{p_1}{2}-\mbf{I}\cdot\mbf{m} \right)
  + 2 \Gamma_\mrm{L}(1-f_\mrm{L}) p_\mrm{T} \, ,
\end{equation}
decouple from the rest. 
This motivates proposal a) for generating 
spatially-separated spin entanglement.
If we prepare the system in an empty state (by tuning the dot levels 
well above the Fermi energy of the lead) and, then, suddenly 
push the dot levels well below the Fermi energy of the left lead,
$-\varepsilon \gg \mrm{k_B}T,\GL$, the double dot will be charged with two
electrons that form a spin singlet, while no triplet component appears.
The time dependence of the singlet generation is illustrated in 
Fig.~\ref{DiscForkPlots}.

Coupling to an external bath,  which flips the spin of an electron or generates a relative phase between the $|\uparrow, \downarrow \rangle$ and $|\downarrow, \uparrow \rangle$ states, induces relaxation from the singlet to the triplet state. 
To model these processes we introduce phenomenologic relaxation rates
$\Gamma_\mrm{S\rightarrow T_0}$, $\Gamma_\mrm{S\rightarrow T_\pm}$, 
$\Gamma_\mrm{T_0\rightarrow S}$, $\Gamma_\mrm{T_\pm\rightarrow S}$,
$\Gamma_\mrm{T_0\rightarrow T_\pm}$, and $\Gamma_\mrm{T_\pm\rightarrow T_0}$. 
To be specific, we choose all of them to be equal, such that we get an 
effective transition rate $\Gamma_\mrm{S\rightarrow T} = 
\Gamma_\mrm{T\rightarrow S}/3 $ which conserves the symmetry between the 
triplets, $p_\mrm{T_\pm} = p_\mrm{T_0}$. 
(A different choice of these parameters does not change the conclusions qualitatively.)
Furthermore, a finite detuning $\Delta \tilde \varepsilon$ and/or finite
asymmetry of the tunnel couplings $\Gamma_\mrm{Lu} \neq \Gamma_\mrm{Ld}$, lead
to a mixture of singlet and triplet states, producing a Werner 
state~\cite{Werner} described by the density matrix
\begin{equation}
  W(F)=F |{\rm S}\rangle\langle {\rm S}| + (1-F) \frac{\openone_4 -|{\rm S}
    \rangle\langle {\rm S}|}{3} \, .
\end{equation}
The parameter $F$ defines the Werner fidelity. Werner states play a crucial role in entanglement purification protocols,~\cite{BennettPRA,Linden} and the
Werner fidelity gives a convenient measure for the possibility to extract
entangled states from a set of Werner states by such protocols.
It has been further shown that for Werner fidelity $1/2 < F \le 1$ there exist
purification protocols to extract states with arbitrary large entanglement
whereas for $F \le 1/2$ the Werner state has to be considered as unentangled. 

Solving the kinetic equations for the reduced density matrix for system a) 
we see that Werner states with fidelity $F = p_\mrm{S}/(p_\mrm{S}+p_\mrm{T}) 
> 1/2$ are accessible also for asymmetric tunneling, detuning and finite
spin-flip relaxation, see Fig.~\ref{DiscForkPlots}.
For weak detuning $\Delta \varepsilon$ the probability to generate a triplet scales with $p_\mrm{T} \approx (\Delta \varepsilon/2\Gamma_\mrm{L})^2$.

To create and detect an enhanced spin-singlet fidelity and to measure the 
relaxation time between singlet and triplet we propose the following scheme 
that is similar to the experiment performed in Ref.~\onlinecite{PettaScience}.
(i) Prepare the system in an empty state.
(ii) Push quickly (i.e. on a time scale faster than both the relaxation times
for the isospin-polarized state and for the singlet-triplet transitions)
the dot levels down well below the Fermi level.
As explained above the double dot will preferably fill up with two electrons 
forming a spin singlet.
(iii) Wait some given time $T$.
As a function of $T$, the imbalance between singlet and triplet decays
exponentially on the time scale given by the relaxation rate, and the Werner
fidelity is reduced, see Fig.~\ref{DiscForkPlots}.

\begin{figure}
\begin{center}
\includegraphics[width=8.2cm]{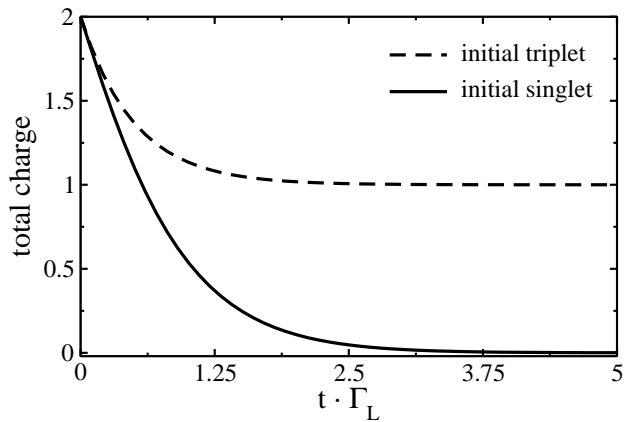}
\end{center}
\caption{\label{SingletTripletDischarge} 
Time evolution of the total charge during the discharging of an initial singlet state compared to an initial triplet state. The system is assumed to be perfectly symmetric, $\Gamma_{\mrm{Lu}} = \Gamma_{\mrm{Ld}}$, $\Delta\varepsilon = 0$. Starting from a singlet the system empties quickly, but it remains singly occupied if we start from a triplet state.}
\end{figure}

To prove that the obtained state, indeed, has an enhanced Werner fidelity,
we analyze how the double dot is depleted.
Depending on whether the initial state is a singlet or triplet, it is 
possible or impossible to extract the two electrons by tunneling to the common
left lead.
This can be seen by realizing that 
$\left( t_{{\rm Lu}} c_{\mrm{u}\sigma} + t_{{\rm Ld}} c_{\mrm{d}\sigma} \right)
\left( t_{{\rm Lu}} c_{\mrm{u}\bar \sigma} + t_{{\rm Ld}} 
c_{\mrm{d}\bar \sigma} \right) \left( 
|\sigma, \bar \sigma \rangle \mp |\bar \sigma, \sigma \rangle \right)
= (1\pm 1) t_{{\rm Lu}} t_{{\rm Ld}} |0\rangle $
is finite for the singlet but vanishes for the triplet state, i.e., only one
of the two electrons forming the triplet can leave.  
As a consequence, the proposed protocol continues in the following way.
(iv) Push the dot levels up well above the Fermi level quickly (again
faster than the relaxation rate for the isospin-polarized state).
(v) Wait some time larger than $1/\GL$ but shorter than the relaxation time of
the isospin-polarized state.
(vi) Measure the total charge on the double dot.
If the charge is zero then the doubly-occupied state was a spin singlet,
whereas if the measured charge is one, it was a triplet.
To illustrate this we show in Fig.~\ref{SingletTripletDischarge} the total
double-dot charge as a function of time for the two cases that the double dot 
initially accommodated a singlet or a triplet, respectively.
The measurement of the total charge on the double dot could be performed by a 
close-by quantum-point contact.
This does not introduce an additional relaxation mechanism for either the
isospin or the singlet and triplet states as the quantum-point contact is
only sensitive to the total charge.

\begin{figure}
\begin{center}
\includegraphics[width=8.2cm]{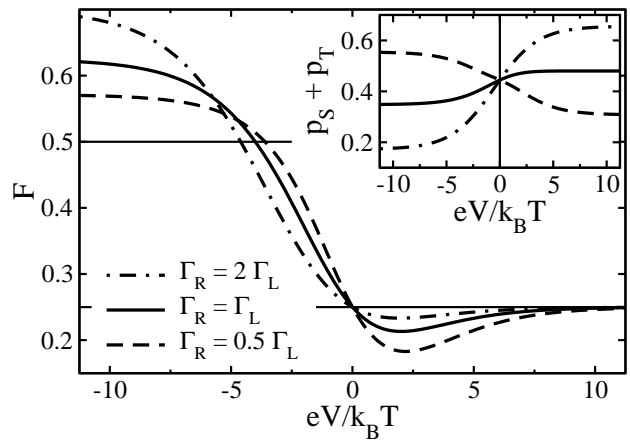}
\end{center}
\caption{\label{ForkPlots} 
The stationary Werner fidelity $F$ vs.\@
bias voltage for $\varepsilon = 0$ and different ratios of the coupling strengths, $\GR/\GL = 2, 1, 0.5$. 
The bias is applied symmetrically, $- \mu_\mrm{L} = \mrm{e}V/2 = \mu_\mrm{R_u} = \mu_\mrm{R_d}$. 
The inset shows the corresponding
  stationary overall probabilities $p_\mrm{S}+p_\mrm{T}$ to find the system doubly occupied.}
\end{figure}

\subsection{Spin Entanglement for Finite Bias Voltage}

Spatially-separated spin entanglement is found also in a steady-state
situation in the setup b) of Fig.~\ref{setups}.
Here we consider the system to be driven out of equilibrium by a bias voltage between the left and the right side.
To keep the discussion transparent we assume in the following symmetric couplings, $\Gamma_\mrm{Lu}=\Gamma_\mrm{Ld}\equiv \Gamma_\mrm{L}$ and $\Gamma_\mrm{R_uu} = \Gamma_\mrm{R_dd}\equiv \Gamma_\mrm{R}$, equal electrochemical potentials in the right leads $\mu_\mrm{R_u}=\mu_\mrm{R_d}$, and vanishing detuning of the dot levels $\Delta \varepsilon =0$.
The leads on the right hand side couple to all isospin components in the same
way.
In a magnetic analogue such a situation corresponds to a dot coupled to one
ferromagnetic and one nonmagnetic lead for which, at large bias voltage, 
spin accumulation occurs.
Similarly, in the present model a finite isospin is accumulated in the 
double dot in the stationary limit.
This, again, leads to an imbalance of singlet and triplet state
probabilities.
The polarity of the bias voltage determines whether the Werner fidelity is
larger or smaller than $1/4$.
If the bias voltage is applied such that the double dot is charged from the
left and decharged to the right lead the isospin polarization is in $+$
direction, and singlets are preferred. 
In this regime the Werner fidelity saturates at $F = (3\,\Gamma_\mrm{L} + 2\,\Gamma_\mrm{R})/(6\,\Gamma_\mrm{L} + 2\,\Gamma_\mrm{R}) $ 
which goes from $1/2$ for $\Gamma_\mrm{L} \gg \Gamma_\mrm{R} $ to $1$
for $\Gamma_\mrm{L} \ll \Gamma_\mrm{R} $.
We have to remark that the fidelity approaches 1 only linearly for 
$\Gamma_\mrm{L} \ll \Gamma_\mrm{R} $, whereas the overall probability to find the double-dot system doubly occupied vanishes quadratically $p_\mrm{S} + p_\mrm{T} \approx 2\,(\Gamma_\mrm{L}/\Gamma_\mrm{R})^{2} $ at the same time.  
If the bias voltage is applied in the opposite direction, triplets are more likely.

\section{Conclusions}

For an experimental realization of our proposal one needs to coherently couple
two quantum dots to a joint reservoir, as has been demonstrated e.g. in 
Ref.~\onlinecite{coherent-coupling}.
The spatial separation of the two dots is only limited by the 
phase-coherence length, which can be several micrometer
in typical semiconductor structures.
The formation of an enhanced spin-singlet fidelity requires tunneling rates 
larger than the spin decoherence time.
Reported values~\cite{PettaScience,koppens} of $T_2^*$ of the order of 10 ns
correspond to a lower limit of $\Gamma$ of the order of $\mu$eV.
For tunnel couplings $\Gamma$ larger than $k_{\rm B}T$ higher-order 
processes such as cotunneling and Kondo-assisted tunneling
become important.
These are neglected in our quantitative analysis but they do not change our 
prediction qualitatively.
In fact, for symmetric tunnel couplings the Hamiltonian acquires a block 
structure and the Hilbert subspace containing the triplet states decouples 
completely from the one for the empty double dot.
In conclusion, the experimental realization of our proposal should be 
feasible by nowadays technology. 

In summary, we proposed two schemes for the generation of spin entanglement 
between two spatially separated electrons in a double-dot system driven out 
of equilibrium.
The underlying mechanism is fundamentally different from those that rely on a 
singlet-triplet energy splitting, where entanglement is generated by a
relaxation of the system to the spin-entangled ground state.
In contrast, we suggest two schemes in which entanglement is a consequence of 
a coherent coupling of two quantum dots to one common lead in combination with
a strong onsite Coulomb interaction to prevent double occupancy of each 
individual dot.
We emphasize that our proposal does not require a finite singlet-triplet 
splitting.
The quick formation of the entangled state on a time scale given by the 
tunneling instead of a singlet-triplet relaxation rate, may be an advantage 
in the context of quantum information processing.

We acknowledge useful discussions with P.W. Brouwer, J. Weis, J. Martinek, Y. Gefen, T. L\"ofwander, and E. Prada. This work was supported by the Landesstiftung Baden-W\"urttemberg via the Kompetenznetz Funktionelle Nanostrukturen, DFG via GRK 726 
and SFB 491, the Swiss SNF and NCCR Nanoscience, as well as by NSF under grant PHY99-07949.

\end{document}